\documentclass[
    ,final            
    ,numbers              
  ]
  {aipproc}

\usepackage{amssymb,amsmath,amstext,amsthm,amsfonts}
\layoutstyle{6x9}

\usepackage[usenames]{color}
\definecolor{magenta}{cmyk}{0,1,0,0}

\begin{document}

\title{Resonance-Parton Duality\\ and the Transverse Response of Nucleons}


\author{Murat Kaskulov}{
  address={Institut fuer Theoretische Physik, Universitaet Giessen, D-35392 Giessen, Germany}
}

\author{Ulrich Mosel}{
  address={Institut fuer Theoretische Physik, Universitaet Giessen, D-35392 Giessen, Germany}
}

\begin{abstract}
QCD-based scaling arguments predict the predominance of longitudinal over transverse electroproduction of pions by terms $\propto Q^2$. However, data from JLAB, Cornell and DESY, covering a wide kinematical range $1 < Q^2 < 11\, {\rm GeV}^2$ and $2\, {\rm GeV} < W < 4\, {\rm GeV}$, do not show this expected behavior. At the same time standard descriptions of pion-electroproduction on nucleons have given a very good description of the longitudinal components of the cross sections. However, these very same models have failed grossly in describing the transverse component. We discuss here a common solution to these two problems by considering the contributions of high-lying ($W > 2$ GeV) nucleon resonances to pion production. The coupling strengths and form factors are obtained through hadron-parton duality. We show that an excellent description of data in a wide range of electron energies and four-momentum transfers can be obtained in such a model.
\end{abstract}

\maketitle


\section{Introduction}
QCD based arguments predict that transverse component of the cross section for the exclusive
reaction $p(e,e'\pi^+)n$ falls off with $1/Q^8$ while the longitudinal component falls with $1/Q^6$  \cite{Brodsky:1974vy,Collins:1996fb}. Exclusive pion production can be used to check this prediction and to find out at in which kinematical regime these simple scaling laws become effective. In the following we discuss these reactions; we draw here on our publications \cite{Kaskulov:2008xc,Kaskulov:2009gp,Kaskulov:2010kf} where further details can be found.

Indeed, at Jefferson Laboratory (JLAB) the exclusive
reaction $p(e,e'\pi^+)n$  has been investigated for a range of photon
virtualities up to $Q^2\simeq 5$~GeV$^2$ at an invariant mass of the $\pi^+n$
system around the onset of deep--inelastic regime,
$W\simeq 2$~GeV~\cite{Horn:2006tm,Horn:2007ug,Tadevosyan:2007yd}.
A separation of the cross section into the transverse $\sigma_{\rm T}$
and longitudinal $\sigma_{\rm L}$ components has been performed.

The data  show that $\sigma_{\rm T}$ is large at JLAB energies \cite{Horn:2007ug}. At $Q^2 = 3.91$~GeV$^2$ $\sigma_{\rm T}$ is by about a factor of two larger than $\sigma_{\rm L}$, constrary to the QCD-based expectations, and at $Q^2=2.15$~GeV$^2$ it has same size as $\sigma_{\rm L}$. Previous measurements at values of $Q^2=1.6~(2.45)$~GeV$^2$~\cite{Horn:2006tm} show a similar problem in the understanding of $\sigma_{\rm T}$. Even at smaller JLAB~\cite{Tadevosyan:2007yd} and much higher Cornell~\cite{Bebek:1977pe} and DESY~\cite{:2007an} values of $Q^2$ there is a disagreement between the simple scaling expectations and experimental data.

The longitudinal cross section $\sigma_{\rm L}$ is well understood in terms of the pion quasi--elastic knockout mechanism~\cite{Neudatchin:2004pu} because of the pion pole at low $-t$ . This makes it possible to study the charge form factor of the pion at momentum transfer much bigger than in the scattering of pions from atomic electrons~\cite{Sullivan:1970yq}. However, the model of Ref.~\cite{Vanderhaeghen:1997ts}, which is based on such a picture and which is generally considered to be a guideline for the experimental analysis and extraction of the pion form factor, underestimates $\sigma_{\rm T}$ at $Q^2=2.15$~GeV$^2$ and at $Q^2=3.91$~GeV$^2$  by about one order of magnitude~\cite{Horn:2007ug}.

\section{Transverse strength in electroproduction}
\begin{figure*}[ht]
\includegraphics[clip=true,width=0.95\columnwidth,angle=0.]{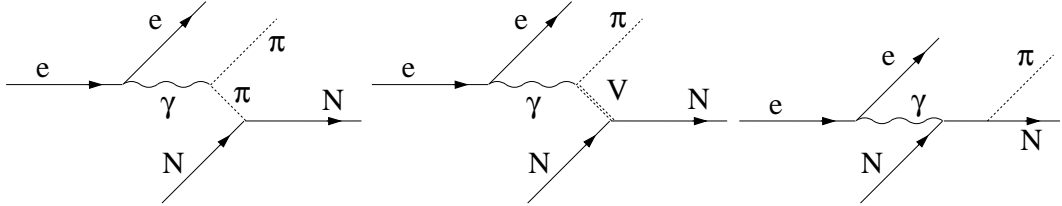}
\caption{
 The diagrams describing
       the hadronic part of the $\pi^+$-- electroproduction amplitude at high
       energies. The leftmost diagram shows the $t$-channel contribution, the middle one the contribution of vector meson exchange and the rightmost one gives the nucleon Born term.      } \label{Figure1}
\end{figure*}

The model of Ref.\ \cite{Vanderhaeghen:1997ts} can be represented by the three amplitudes shown in Fig.\ \ref{Figure1}.
The leftmost Reggeized $t$-channel term dominates the longitudinal strength which thus always dominates at forward momenta, with relatively small contributions from vector meson exchange (middle graph). The right graph, involving a nucleon-pole $s$-channel diagram, is necessary for gauge invariance; its main, but small, contribution is to the transverse cross section. At the kinematics of the relevant electroproduction experiments the energies are high enough to excite also nucleon resonances; the invariant masses of the $(\gamma^*p)$ system are between approximately 2 and 4 GeV. We identify these high-lying resonances with partonic excitations leading to DIS \cite{Domokos:1971ds} and invoke the continuity in going from an inclusive final DIS state to the exclusive pion production \cite{Bjorken:1973gc}. During this transition the transverse strength of DIS remains intact. We thus expect that the inclusion of these resonance excitations into the $s$-channel diagram enhances the transverse scattering while leaving the longitudinal strength originating in the $t$-channel diagram intact. Adding such a resonance (DIS) component to the $t-$\emph{channel + Born model} of \cite{Vanderhaeghen:1997ts} constitutes the central point of our theory which also contains an improved treatment of gauge invariance accounting for the difference in the electromagnetic formfactors for pions and protons \cite{Kaskulov:2008xc,Kaskulov:2009gp,Kaskulov:2010kf}.

The natural point to add the resonance contributions is in the rightmost diagram of Fig.\ \ref{Figure1} that depicts the so-called Born graph, i.e.\ a $s$-channel graph with an intermediate nucleon in the original theory of Ref.~\cite{Vanderhaeghen:1997ts}. The resonance excitations are also $s$-channel contributions which -- because of their special coupling -- are gauge invariant by themselves \cite{Penner:2002md}.

In the experiments quoted the invariant masses of the $(\gamma^*,N)$ system are all $W > 2$ GeV, i.e.\ they all lie above the region of well-established, separated nucleon resonances. Thus, the Born term becomes a sum over individual resonances that can be replaced by an integral over resonances with average coupling constants and form factors 
\begin{eqnarray}      \label{sumres}
B = \sum \limits_{i} r(M_i) c(M_i)
\frac{F(Q^2,M_i^2)}{s-M_i^2+i0^+} &\Rightarrow&
\int \limits_{M_p^2}^{\infty} dM_i^2 \rho(M_i^2) r(M_i^2)c(M_i^2)
\frac{F(Q^2,M_i^2)}{s-M_i^2+i0^+} ~,
\end{eqnarray}
where $r(M_i)$ and $c(M_i)$ are the electromagnetic and strong couplings, respectively, and $F(Q^2,M_i^2)$ is the electromagnetic form factor. Here $\rho(M_i^2)$ is the density of resonances with mass $M_i$. So far unknown are here the couplings $r$ and $c$ and the form factors $F$ in Eq.\ (\ref{sumres}).

Our aim is to maintain the transverse character of DIS, which follows from a parton picture, when going to the exclusive limit of a resonance decay. We thus have to establish a connection between these two pictures. Bloom and Gilman \cite{Bloom:1970xb,Bloom:1971ye} have shown that the total DIS strength follows closely the average behavior of nucleon resonances (for a more recent review see also \cite{Melnitchouk:2005zr}). We, therefore, now use this Bloom and Gilman duality in its local form
\begin{equation}
\label{BG}
F_2^p(x_{\rm B},Q^2) = \sum_{i} (M_i^2-M_p^2+Q^2)  W(Q^2,M_i) \delta(s - M_i^2),
\end{equation}
where $x_{\rm B}$ stands for the Bjorken scaling variable and the deep inelastic
structure function $F_2^p(x_{\rm B},Q^2)$ is expressed as a sum of resonances. 
$W(Q^2,M_i)$ defines the $i$th resonance contribution to the $\gamma^* p$ forward
scattering amplitude; it is essentially the electromagnetic coupling constant
$r(M_i)$ times the resonance form factor $F(Q^2,M_i)$.
Eq.\ (\ref{BG}) links the partonic content of nucleon resonances with their hadronic structure.
Since the density of resonances in Eq.\ (\ref{sumres}) is a steeply increasing function of invariant mass, it follows from Eq.\ (\ref{BG}) that the electromagnetic coupling to nucleon resonances must be decreasing with mass since $F_2$ is finite. This leads to a natural cut-off for the number of resonances and makes the integral in Eq.\ (\ref{sumres}) finite.

\begin{figure}[ht]
\includegraphics[clip=true,width=0.50\columnwidth,angle=0.]
{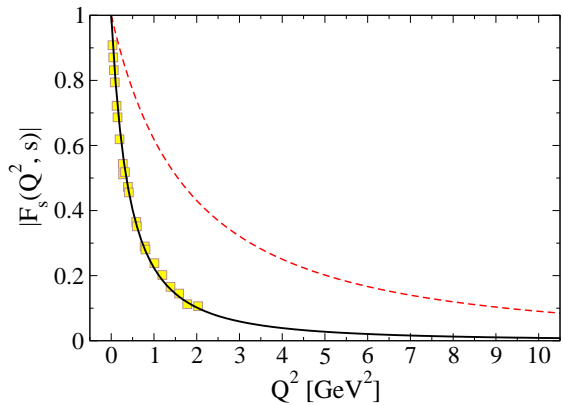}
\caption{
\label{F1onoff}
(Color online) The $Q^2$ dependence of the absolute value
of the transition form factor $|F_s(Q^2,s)|$ in Eq.~\ref{sumres} (dashed curve) at $\sqrt{s}=2.2$~GeV. The solid curve
describes the proton Dirac form factor in
comparison with data. From \cite{Kaskulov:2010kf}.
}
\end{figure}

In a further step the combined nucleon-resonance contribution to the s-channel is absorbed into an effective Born-term with a nucleon pole only, but a modified form factor $F_s$ which now contains all the effects of the resonances
\begin{equation}
B = \frac{F_s(Q^2,s)}{s-M_p^ 2+i0^+} ~.
\end{equation}
 Parametrizing the form factor in a dipole form leads to the conclusion that the cut-off must increase with the mass of the resonance \cite{Elitzur:1971tg}. We, therefore, expect that the effective form factor $F_s$ is considerably harder than that of the nucleon alone. This is indeed born out by the calculations as can be seen in Fig.\ \ref{F1onoff}.

\begin{figure*}[htb]
\includegraphics[clip=true,width=0.8\columnwidth,angle=0.]{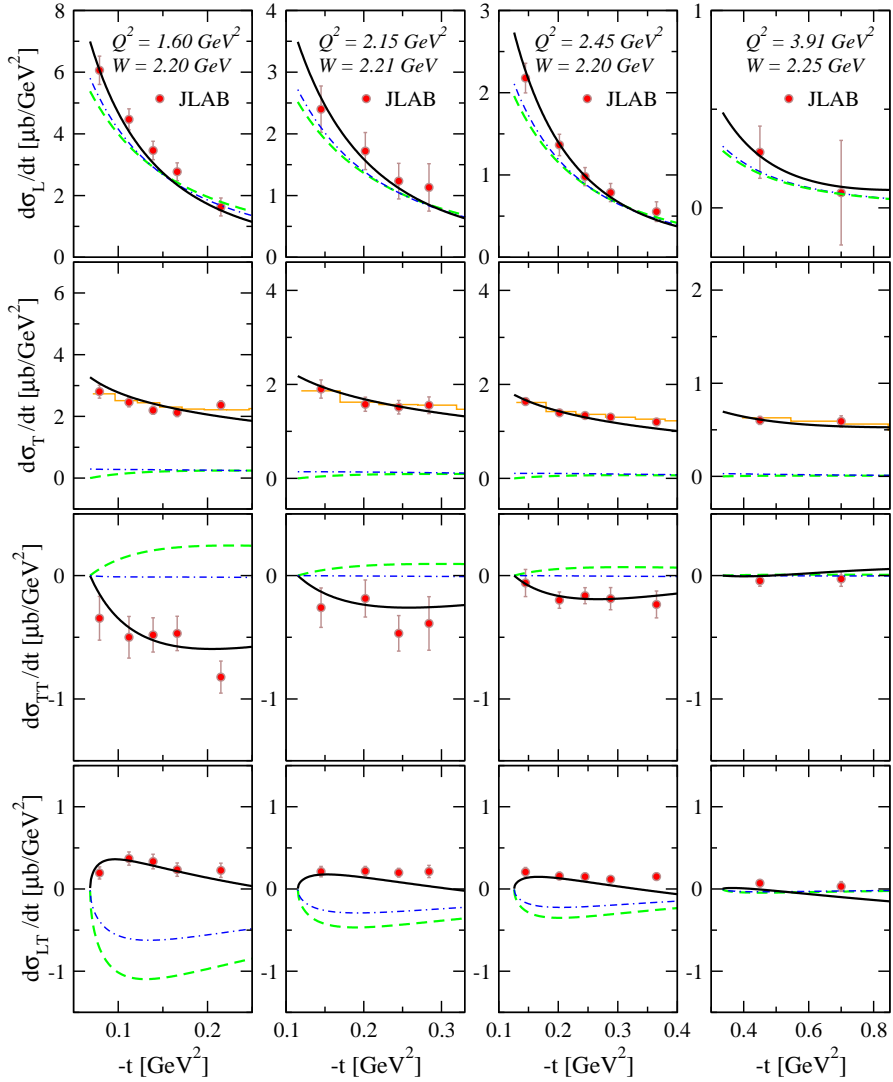}
\caption{\label{Horn1}
 (Color online) $-t$ dependence of \textsc{l/t} partial transverse
$d\sigma_{\rm T}/dt$, longitudinal $d\sigma_{\rm L}/dt$ and interference
$d\sigma_{\rm TT}/dt$ and $d\sigma_{\rm LT}/dt$ differential cross sections
in exclusive reaction $p(\gamma^*,\pi^+)n$. The experimental data are from
the $F\pi$-2~\cite{Horn:2006tm} and $\pi$-CT~\cite{Horn:2007ug}
experiments at JLAB. The numbers displayed in the plots are the average $(Q^2,W)$ values.
The dashed curves correspond to the exchange of the $\pi$-Regge trajectory alone.
The dash-dotted curves are obtained with the on-mass-shell form
factors in the nucleon-pole contribution and exchange of the
$\rho(770)/a_2(1320)$-trajectory. The solid curves describe the model results with the
resonance contributions. The data points in each $(Q^2,W)$ bin correspond
to slightly different values of $Q^2$ and $W$ for the various $-t$ bins.
The calculations are performed for values of $Q^2$ and $W$ corresponding to
the first $-t$ bin. The histograms for $d\sigma_{\rm T}/dt$ are the results
from~\cite{Kaskulov:2008xc}. From \cite{Kaskulov:2010kf}.
}
\end{figure*}

Fig.\ \ref{Horn1} shows the results of the calculations in comparison with the data of Ref.\ \cite{Horn:2006tm,Horn:2007ug} for all four cross sections. The solid lines that represent the results of our calculations describe all cross sections, the longitudinal, the transverse and the interference ones, very well. The transverse strength is nearly entirely built up by the resonance contributions that are considerably bigger and fall off much more weakly with $Q^2$ than the ones obtained from the nucleon Born term alone (see Fig.~\ref{F1onoff}).  The resonances also contribute about 30\% to the longitudinal cross section at forward angles (small $-t$) where the major contribution comes from the $t$-channel graph with the nucleon Born term alone. For the interference cross sections $\sigma_{\rm TT}$ and $\sigma_{\rm LT}$ the sign even changes when the resonances are taken into account.

\begin{figure*}[ht]
\includegraphics[clip=true,width=0.8\columnwidth,angle=0.]
{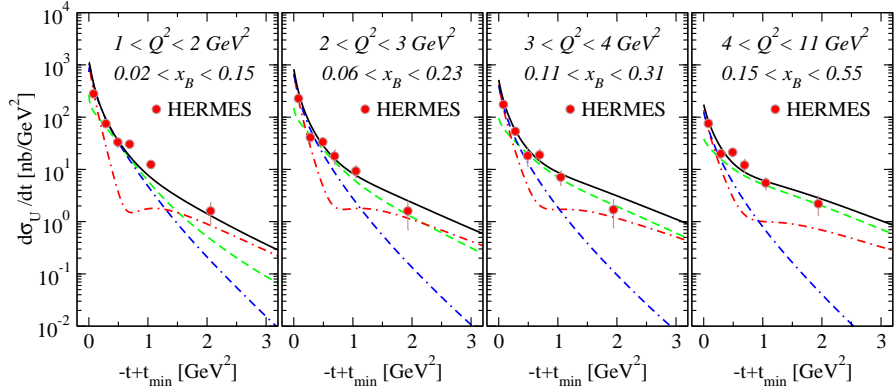}
\caption{\label{EffHermes} (Color online) $-t+t_{min}$ dependence of the
differential cross section $d\sigma_U/dt = d\sigma_{\rm T}/dt + \epsilon d\sigma_{\rm
L}/dt$ in exclusive reaction $p(\gamma^*,\pi^+)n$ at HERMES.
The experimental data are from Ref.~\cite{:2007an}.
The calculations are performed for the average values of
$(Q^2,x_{\rm B})$ in a given $Q^2$ and Bjorken $x_{\rm B}$ bin.
The solid curves are the full model results.
The dash-dotted curves correspond to the longitudinal $\epsilon d\sigma_{\rm L}/dt$
and the dashed curves to the transverse $d\sigma_{\rm T}/dt$ components of the
cross section.
The dash-dash-dotted curves describe the results without the
resonance/partonic effects. From \cite{Kaskulov:2010kf}.}
\vspace{-0.5cm}
\end{figure*}

In Fig.\ \ref{EffHermes} we show that this very same picture also works remarkably well at the much higher momentum transfers $Q^2$ and invariant masses $W$ reached in the HERMES experiment \cite{:2007an}. Even in the kinematical windows $4 < Q^2 < 11$ GeV$^2$ the difference between the dash-dash-dotted curve (no resonance contributions) and the dashed curve (resonances included) shows the dramatic impact of the resonance contributions which are essential in describing the cross section at larger $-t$.

As the HERMES kinematics are quite close to those expected for JLAB at 12 GeV we predict that again high-lying resonances determine the transverse cross sections at larger $-t$. Fig.\ 23 in Ref.\ \cite{Kaskulov:2010kf} detailed predictions for the $L/T$ separated cross sections are given both for $\pi^+$ and $\pi^-$ production. In particular we predict that $\pi^-$ production is largely longitudinal.

This model has recently been extended  to the photo- and electroproduction of $\pi^0$ \cite{Kaskulov:2011wd}; again a very good description of all available data is reached without any new parameters. The transition from photoproduction ($Q^2 = 0$) to electroproduction shows that the resonance contributions assume a larger and larger role with increasing $Q^2$. While at the photon point they just fill in the diffractive dip in the otherwise Regge-dominated cross section at higher $Q^2$ they become more and more dominant.

\section{Summary}
The large transverse strength observed in various experiments on exclusive pion production has been a long-standing puzzle. We have resolved this open problem by adding to the usual $t$\emph{-channel + Born} term description the contribution of high-lying nucleon resonances as an effective description of DIS excitations. By using quark-hadron duality we have been able to link the properties of these resonances to the partonic degrees of freedom. We thus treat this contribution to the exclusive production as the limiting case of inclusive DIS processes that are predominantly transverse. Such a model, that makes use only of quite general average properties of nucleon resonances, is able to describe all the available exclusive electroproduction data for pions. In particular the transverse response of nucleons is determined by their resonance/parton excitations.

\begin{theacknowledgments}
This work has been supported by DFG through the TR16 and by BMBF.
\end{theacknowledgments}

\end{document}